\documentclass[%
preprint,
amsmath,amssymb,
aps,
pre,
]{revtex4-2}

\usepackage{graphicx}
\usepackage{dcolumn}
\usepackage{bm}

\usepackage{times}
\usepackage[plain]{fullpage}
\usepackage{graphicx}
\usepackage{amsmath}
\usepackage{changes}
\usepackage[caption=false]{subfig}
\colorlet{Changes@Color}{red}

\begin{document}




\title{Weakly nonlinear ion sound waves in gravitational systems}

\author{P.\ Guio$^1$ and   H.~L.\ P{\'e}cseli$^2$}
\affiliation{ $^1$Department of Physics and Astronomy,
University College London\\
Gower Street,
London WC1E 6BT,
United Kingdom\\
$^2$University of Oslo, Physics Department,
Box 1048 Blindern, N-0316 Oslo, Norway}
\date{\today}




\begin{abstract}
Ion sound waves are studied in a plasma subject to gravitational field.
Such systems are interesting by exhibiting a wave growth
that is a result of energy flux conservation in inhomogeneous systems.
The increasing wave amplitude gives rise to an enhanced interaction 
between waves and plasma particles that can be modeled by a 
modified Korteweg-de Vries equation. Analytical results are compared
with numerical Particle-in-Cell simulations of the problem. Our
code assumes isothermally Boltzmann distributed electrons while
the ion component is treated as a collection of
individual  particles interacting through collective electric fields. Deviations from quasi neutrality are allowed for.
\end{abstract}

\maketitle 
\section{Steady state}

We consider a hot plasma in a gravitational field in the vertical $z$-direction, with gravity pointing in the downwards direction.
Steady state static solutions with $\overline{u}_z = 0$ are readily 
obtained for the case where we have a balance between gravitational 
effects and thermal particle pressures. For this case $\overline{\phi} =
-z M g/e$ and the steady state vertical electric field is constant,
$\overline{\bf E} = \widehat{\bf z}Mg/e$ in the
positive $z$-direction so that the constant gravitational force
is balanced by the ambipolar electric field induced by the charge separation
caused by the finite electron pressure. This is incidentally an
interesting result: with a constant electric field we have here
the right hand side of Poisson's equation $\nabla\cdot {\bf E}
= e (n_i-n_e)/\varepsilon_0$ to vanish identically, so that the 
steady state solution
is quasi-neutral, $n_e = n_i$, even though no assumption of
quasi neutrality was made explicitly.
For the plasma density we find
   \begin{equation}
   \overline{n}(z)=n_0\exp\left(-zg/C_s^2\right) ,  
\label{vert_grad}
   \end{equation}
with $C_s\equiv\sqrt{T_e/M}$ being the ion sound speed, here for 
cold ions and warm electrons. We can introduce
a vertical scale length $L_{gc} \equiv C_s^2/g$. Temperatures are in energy units, i.e., without Boltzmann's constant.

More generally
both ions and electrons will contribute.
The classical and simplest of these 
equilibrium solutions \cite{pannekoek_1922,rosseland_1924} 
is found for isothermal 
conditions $T_e = T_i \equiv T$,
with the plasma density varying as 
$\overline{n}=n_0\exp(-\frac{1}{2}z(m+M)g/T)$. In this case
the constant gravitational acceleration we have $\overline{n}(m+M)g$ to balance
the plasma pressure $2Td\overline{n}/dz$. The
effect of gravity on the electrons is negligible, but they respond to
the collective electric fields.
For steady state solutions we can assume both electrons and ions to be
in an isothermal Boltzmann equilibrium, possibly with different temperatures,
i.e.\ 
$$
n_e = n_{0e}\exp\left(\frac{e\phi}{T_e}\right)
\hspace{5mm}\mbox{and}\hspace{5mm}
n_i = n_{0i}\exp\left(\frac{-e\phi-Mgz}{T_i}\right),
$$
where we ignored the effect of a constant gravitational force
on the electrons.
We can impose neutrality at the position where $\phi = 0$, 
taken to be $z=0$, to give 
$n_{0e} = n_{0i}\equiv n_0$. To determine the electrostatic potential we
can then insert into Poisson's equation $\nabla^2\phi
= e (n_e-n_i)/\varepsilon_0$ to give $e\phi =
-MgzT_e/(T_i+T_e)$, $E=g(M/e)T_e/(T_i+T_e) = $~const. and
$n_e=n_0\exp(-z M g/(T_i+T_e))$, $n_i = n_0\exp\left(-z M g/(T_e+T_i)\right)$, i.e.\
$n_i=n_e$ also for $T_i\neq 0$.  The present results contain
the Rosseland-Pannekoek isothermal equilibrium 
\cite{pannekoek_1922,rosseland_1924} as a special limit.
In principle, the results is correct for any intensity of the 
gravitational field. 

The steady state solution outlined here assumes one ion species only.
If we insert another singly charged
lighter ion species the gravitational force is smaller
on this, while the force from the vertical electric field is the same.
This lighter species will consequently be accelerated in the
vertical direction to give the ``polar wind'' \cite{ganguli_1996}.
In the present study we will discuss other forms of acceleration
and restrict the analysis to one ion species. The problem addresses
vertical ion flows in a gravitational field and can therefore be
analyzed in one spatial dimension.

\section{Linear wave propagation}

This section summarizes the properties of linear wave propagation. As
a reference case we include also a summary for low frequency
waves propagating in a homogeneous magnetized plasma.

\subsection{Homogeneous magnetized plasma conditions}

\begin{figure}
\includegraphics[width=0.45\columnwidth]{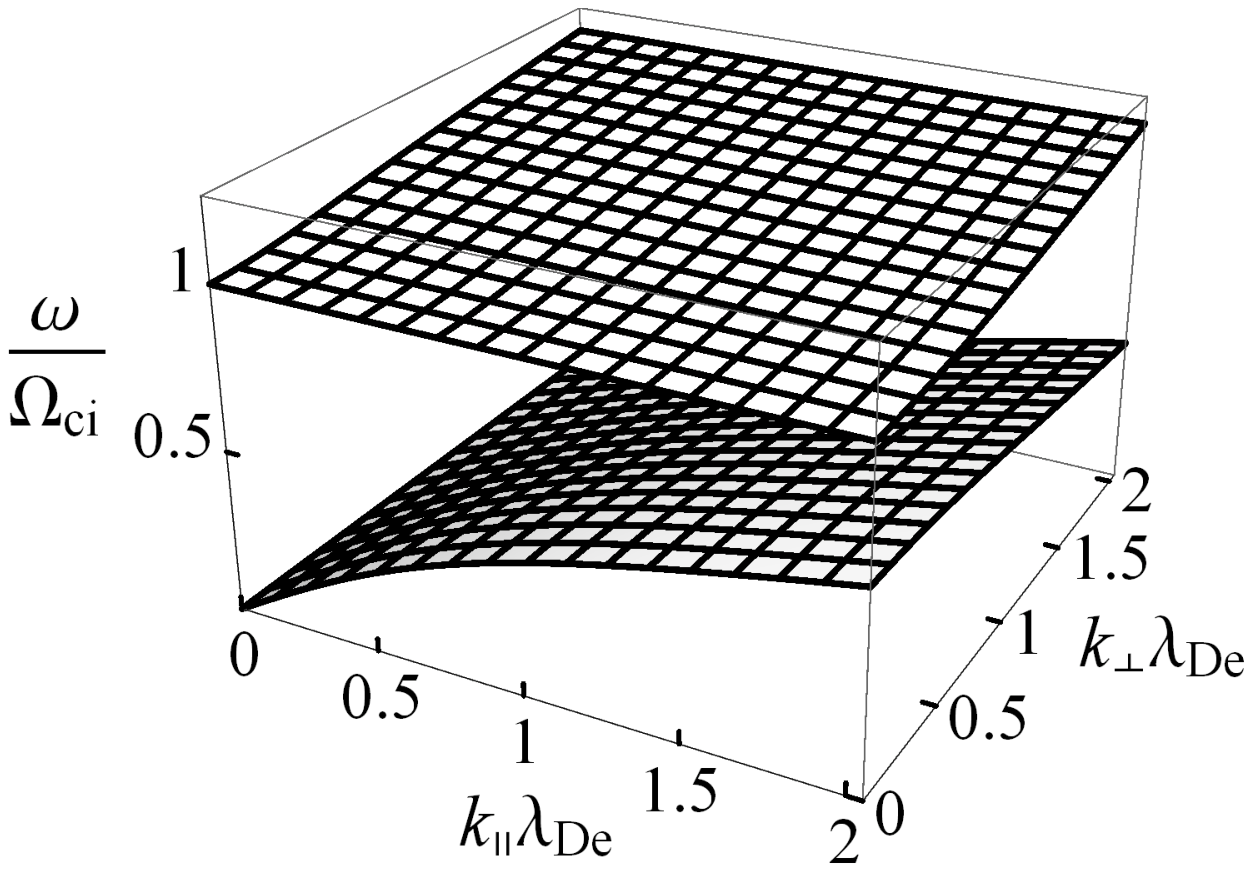}\\
\vspace{5mm}
\includegraphics[width=0.33\columnwidth]{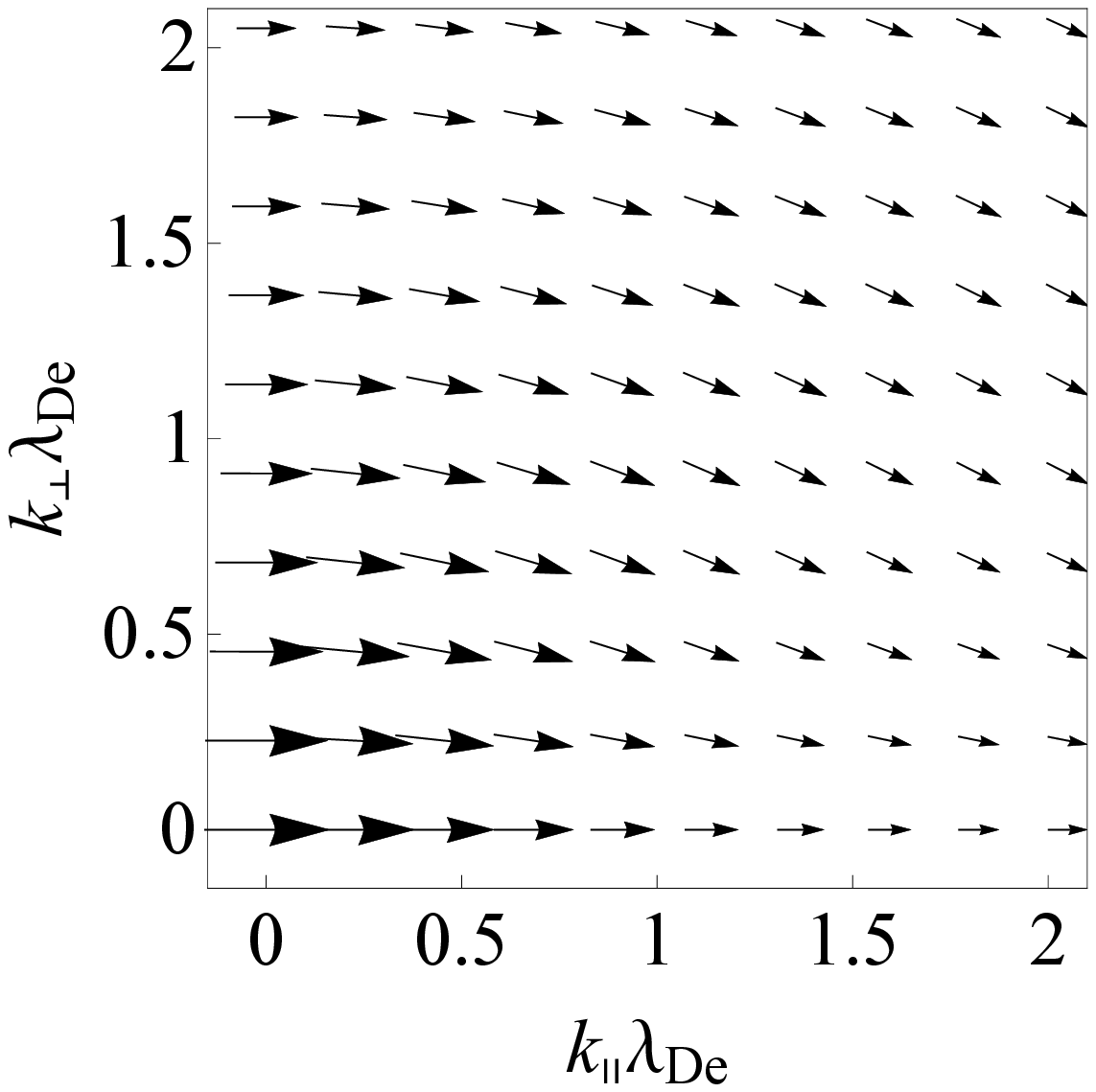}
\caption{\it Linear dispersion relation for ion waves propagating in a
homogeneous magnetized plasma with $\Omega_{ci} > \Omega_{pi}$. 
There are two branches: a low frequency branch $\omega < \Omega_{pi}$
relevant here, and a high frequency wave component $\omega\approx\Omega_{ci}$.
The variation of the group velocity vectors for the
low frequency branch are shown as well. We have $\Omega_{ci} = 2\, \Omega_{pi}$.
and $T_e = 10\, T_i$.}  
\label{fig:lindisp}
\end{figure}

For homogeneous magnetized plasma conditions the linear dispersion relation
$\omega = \omega({\bf k}$ can be found in the literature \cite{pecseli_2012}.
Two limiting cases can be recognized: $\Omega_{ci} > \Omega_{pi}$ and 
$\Omega_{ci} < \Omega_{pi}$ in terms of ion cyclotron and ion plasma frequencies.
A previous study \cite{guio_pecseli_2016} discussed weakly nonlinear 
ion waves for $\Omega_{ci} < \Omega_{pi}$. The other limit will
be relevancy for the present analysis. The linear dispersion relation 
and the variation of the group velocity vectors is
shown in Fig.~\ref{fig:lindisp}.

We find that the group velocity vectors are nearly parallel to ${\bf B}$
for the low frequency branch. A localized perturbation will therefore propagate 
along magnetic field lines with
small dispersion in the direction $\perp{\bf B}$ for this wavetype.
The waveforms analyzed in the following belong to the low
frequency branch. A spatially one dimensional study is justified by
considering conditions where a waveguide mode excited in a
magnetic flux tube with enhanced electron temperatures, $T_\gg T_i$
compared to the surrounding plasma similar to a previous study
\cite{guio_pecseli_2016}.

\subsection{Inhomogenous plasma conditions with gravity}

Propagation of waves in a gravitational field in a horizontally
striated environment has an equivalent in the neutral atmosphere
\cite{hines_1960} where a vertical density gradient is found, and the problem has similarities with the one considered in the present study. Here we use the linearized ion continuity equation and momentum equations first for cold ions for illustration.
Introduce the potential as $\phi=\overline{\phi}
+\widetilde{\phi}$ and $n=\overline{n}(z)
+\widetilde{n}$ to separate the fluctuating parts from the 
steady state equilibrium values. With the present assumptions,
the velocity $u_z$ has fluctuating components only so $^{\widetilde{}}\,$ is
omitted here.
Assume also Boltzmann distributed electrons,
$n_e = \overline{n}_e(z)
+\widetilde{n}_e=n_0\exp(e\phi/T_e) = n_0\exp(e(\overline{\phi}
+\widetilde{\phi})/T_e)$, and quasi-neutrality,
$n_e\approx n_i\equiv n$.  The reference density
$n_0$ is found where the potential $\overline{\phi}$ vanishes
at steady state and corresponds to $\overline{n}_e(z=0)$.
Linearizing the electron equation we have 
$\widetilde{n}_e = n_0(e\widetilde{\phi}/T_e)\exp(e\overline{\phi}/T_e)
\equiv \overline{n}(z)e\widetilde{\phi}/T_e$
giving the linear ion continuity equation in the form
   \begin{equation}
\frac{e}{T_e}\frac{\partial}{\partial t}\widetilde{\phi}  =
-\frac{\partial u_z}{\partial z}+\frac{u_z}{L_{gc}},
\label{vert_prop1}
   \end{equation}
and the linear ion momentum equation for cold ions becomes
   \begin{equation}
\frac{\partial}{\partial t}u_z  =
-\frac{e}{M}\frac{\partial \widetilde{\phi}}{\partial z}.
\label{vert_prop2}
   \end{equation}

Eliminating $\widetilde{\phi}$ we find for a plane wave solution
$\exp(-i(\omega t-k_z z))$ a complex dispersion relation in the form
   \begin{equation}
\omega^2 -i g k_z  -C_s^2 k_z^2=0 ,
\label{simp_eq}
   \end{equation}
where $g$ is the gravitational acceleration, here taken constant.
If we assume an initial perturbation with real $k$ we find a
complex frequency
$$
\omega = \pm\sqrt{i g k_z +C_s^2 k_z^2}
$$
The interesting feature is that plane 
waves propagating in the positive $z$-direction
appear to be unstable, while waves propagating in 
the opposite direction are damped. As stated, this refers to a plane 
wave excited initially.
It is here even more interesting to have a wave excited at a boundary
say at $z = 0$ with a real frequency $\omega$, 
and investigate its spatial variation. For this problem we have
from (\ref{simp_eq}) the result
   \begin{equation}
k_z = -\frac{1}{2C_s^2}\left(ig\pm\sqrt{4 C_s^2\omega^2-g^2}\right).
\label{spatial_simp_eq}
   \end{equation}
The spatial variation of for instance the 
fluctuating linear ion fluid velocity
will be given by $\exp\big(-i(\omega t -k_z z)\big)$, or
\begin{eqnarray}
u_z(z,t)&=&U_0\exp\left(\frac{1}{2}z g/C_s^2\right)\nonumber\\
&&\hspace{-20mm}
\times\exp\left(\pm i \frac{z}{2C_s^2} \sqrt{4C_s^2\omega^2-g^2}\right)
\exp(-i\omega t),
\label{uz}
\end{eqnarray}
showing that the wave increases in amplitude as it propagates 
upwards in the vertical direction for $z > 0$. For downwards direction of propagation, $z < 0$  we find a wave damping. Note the cut-off at
$\omega_c = \frac{1}{2}g/C_s$. For real $\omega$ and complex $k$
we have no wave propagation for $\omega < \omega_c$.

A physical argument for the observed wave growth can be given by 
considering the lowest order contribution to the kinetic 
wave energy density $\frac{1}{2}\overline{n}M u_z^2$. The 
wave energy density flux is then 
to the same accuracy $\frac{1}{2}\overline{n}M u_z^2C_s$
with a constant $C_s$ for the given conditions. Since 
$\overline{n}\rightarrow 0$ for $z\rightarrow\infty$ we must 
at the same time have
$u_z^2\rightarrow\infty$ to keep the flux constant. 
The time averaged wave energy density flux is for
$\omega\gg \omega_c$ given as $\frac{1}{2}\overline{n}M |u_z|^2 C_s\approx
\frac{1}{2}M C_s U_0^2 =$~constant since the $z$-variation from 
$\overline{n}$ cancels the $z$-variation from $|u_z|^2$, as expected.
The analysis of the potential energy associated with the wave can be 
analyzed in the same manner. The argument cannot readily be applied to
the initial value problem: if we in that case take a plane wave at
$t=0$, the initial wave energy density will become inhomogeneously
distributed.

\section{Consequences of finite ion temperatures and 
deviations from quasi-neutrality}

\subsection{Finite ion temperatures}

A finite ion temperature
changes the isothermal steady state solution to
$T_i\ln \overline{n}(z) = - e\overline{\phi}(z) - Mgz$ for the ions and
$T_e\ln \overline{n}(z) = e \overline{\phi}(z)$ for the electrons
so that 
\begin{equation}
\overline{n}(z) = n_0\exp \left(-z\frac{Mg}{T_e+T_i}\right),
\label{iontemp-gravity}
\end{equation}
and $e\overline{\phi}(z) = -z M g T_e/(T_e+T_i)$,
giving a modified expression for the steady state vertical length scale
$L_{gc} = (T_e+T_i)/(M g)$.

The basic equations are as follows. 
Introducing $\eta\equiv \widetilde{n}_i/\overline{n}$, 
the linearized ion continuity equation is
$$
\frac{\partial\eta}{\partial t}+
{u}_z\frac{d\ln \overline{n}}{d z}
+\frac{\partial u_z}{\partial z} =0,
$$ 
where ${d\ln \overline{n}}/{d z}=-1/L_{gc}$.

With $p_i = \overline{p}_i(z)+ \widetilde{p}_i$,
$\phi = \overline{\phi}(z)+ \widetilde{\phi}$,
$n = \overline{n}(z)+ \widetilde{n}$, etc. we can write
the ion momentum equation as
$$
M\frac{D u_z}{D t} =
-\frac{1}{\overline{n}(z)+ \widetilde{n}}\frac{\partial}{\partial z}
\left(\overline{p}_i(z)+ \widetilde{p}_i\right)
-e\frac{\partial }{\partial z}
\left(\overline{\phi}(z)+ \widetilde{\phi}\right)-Mg.
$$
Ignoring products of small terms we find
\begin{eqnarray*}
M\frac{\partial u_z}{\partial t}&=&
-\frac{1}{\overline{n}(z)}
\left(1- \frac{\widetilde{n}}{\overline{n}(z)}\right)
\frac{\partial}{\partial z}
\left(\overline{p}_i(z)+ \widetilde{p}_i\right)
-e\frac{\partial }{\partial z}
\left(\overline{\phi}(z)+ \widetilde{\phi}\right)-Mg\\
&=& -\frac{1}{\overline{n}(z)}\frac{\partial}{\partial z}
\overline{p}_i(z)
+\frac{\widetilde{n}}{\overline{n}^2(z)}\frac{\partial}{\partial z}
\overline{p}_i(z)-\frac{1}{\overline{n}(z)}\frac{\partial}{\partial z}
\widetilde{p}_i
-e\frac{\partial }{\partial z}
\left(\overline{\phi}(z)+ \widetilde{\phi}\right)-Mg\\
&=& 
\frac{\widetilde{n}}{\overline{n}^2(z)}\frac{\partial}{\partial z}
\overline{p}_i(z)-\frac{1}{\overline{n}(z)}\frac{\partial}{\partial z}
\widetilde{p}_i
-e\frac{\partial }{\partial z}\widetilde{\phi}.
\end{eqnarray*}
We used
\begin{equation}
-\frac{1}{\overline{n}(z)}\frac{\partial}{\partial z}
\overline{p}_i(z)
-e\frac{\partial }{\partial z}\overline{\phi}(z)-Mg = 0,
\label{steady-state}
\end{equation}
due to  the assumed isothermal steady state condition. 
We took the ion dynamics to
be adiabatic with $\gamma = C_P / C_V$ 
being the ratio of specific heats.  It is readily demonstrated that
(\ref{steady-state}) is consistent 
with the assumed isothermal condition
for the ion component in steady state, 
giving $\overline{p}_i(z) = \overline{n}(z) T_i$ .

The electron component is also
here assumed to be a Boltzmann distribution at all times with constant 
temperature $T_e$, i.e.\ $n_e=n_0\exp(e\phi/ T_e)$
we linearize this expression as
$$
n_e\equiv \overline{n}+\widetilde{n}_e= 
n_0\exp\left(\frac{e\overline{\phi}+e\widetilde{\phi}}{ T_e}\right)
\approx n_0\exp\left(\frac{e\overline{\phi}}{ T_e}\right)
\left(1+ \frac{e\widetilde{\phi}}{ T_e}\right)
$$
This result gives $\widetilde{n}_e= (e\phi/ T_e)
n_0\exp({e\overline{\phi}}/{ T_e})$, or
$e\phi/ T_e=\eta_e$.

We  use 
$n = n_0(p/p_0)^{1/\gamma}$ where $p=n T_i$ to 
obtain a dynamic equation for the ion temperature. 
This inserted into the equation of ion continuity
gives after some simple manipulations the ion pressure equation
\begin{equation}
 \left(\frac{\partial}{\partial t}+
u_z\frac{\partial}{\partial z}\right)p= 
-\gamma p\frac{\partial}{\partial z}{u}_z   ,
\label{ion_pressure_eq}
\end{equation}
where the consequences of compressibility  appear
explicitly by the right hand side. The spatial derivative-terms on the left side account for the convection of pressure perturbations.

Linearizing the ion pressure equation we have
$$
\frac{\partial \widetilde{p}_i}{\partial t}+
u_z\frac{d\overline{p}_i}{d z}=
-\gamma  \overline{p}_i\frac{\partial}{\partial z}{u}_z .
$$
Introducing the normalized quantity
$\zeta\equiv \widetilde{p}_i/ \overline{p}_i$ we find
$$
\frac{\partial \zeta}{\partial t}+
u_z\frac{d \ln\overline{p}_i}{d z}=
-\gamma  \frac{\partial}{\partial z}{u}_z .
$$

We use
$$
\frac{\partial \widetilde{p}_i}{\partial z}\equiv
\frac{\partial {\zeta}\,\overline{p}_i}{\partial z}=
\overline{p}_i\,\frac{\partial {\zeta}}{\partial z}+
{\zeta}\,\frac{d \overline{p}_i}{d z},
$$
and with $\overline{p}_i= \overline{n} T_i$ 
find by the linearized ion momentum equation
$$
M\frac{\partial u_z}{\partial t}=
(\eta-\zeta )T_i\frac{d \ln\overline{n}(z)}{d z}
-T_i\frac{\partial\zeta}{\partial z}
-e\frac{\partial }{\partial z}\widetilde{\phi}.
$$

\subsection{Dispersion: Poisson's equation}

With Boltzmann distributed electrons, 
Poisson's equation has the form
   \begin{equation}
\frac{\partial^2 {\phi}}{\partial z^2}=
\frac{e}{\varepsilon_0}\left(n_e-n_i\right)=
\frac{e}{\varepsilon_0}\left(\overline{n}
\exp(e{\phi}/ T_e)-n_i\right).
\label{poisson}
   \end{equation}
With the present approximations,
this equation is the only one where $T_e$ appears.
Linearizing (\ref{poisson}) we find
   \begin{eqnarray}
\frac{\partial^2 \widetilde{\phi}}{\partial z^2}&=&
\frac{e}{\varepsilon_0}\left(n_0
\exp\left(\frac{e\overline{\phi}(z)}{ T_e}\right)
\frac{e\widetilde{\phi}}{ T_e}
-\widetilde{n}_i\right)=
\frac{e}{\varepsilon_0}
\left(\overline{n}(z)\frac{e\widetilde{\phi}}{ T_e}-
\widetilde{n}_i\right)
\nonumber\\
\frac{\partial^2 e\widetilde{\phi}/ T_e}{\partial z^2}
&=& \frac{e^2\overline{n}(z)}{\varepsilon_0 T_e}
\left(\frac{e\widetilde{\phi}}{ T_e}-\eta\right).
\label{lin-poisson}
   \end{eqnarray}
The latter form contains the Debye length explicitly on the right hand side. For the present problem
we have $\lambda_{De}=\sqrt{\varepsilon_0T_e/(e^2\overline{n}(z))}$. As $z\rightarrow \infty$ we have
$\lambda_{De}(z)\rightarrow \infty$ and (\ref{lin-poisson}) shows that
the assumption of quasi neutrality will necessarily break down above some
altitude for any initial condition characterized by some given wavelength.

The complete set of linear equation for the normalized quantities
$\eta=\widetilde{n}_i/\overline{n}$ and 
$\zeta=\widetilde{p}_i/\overline{p}_i$ is
   \begin{eqnarray}
\frac{\partial\eta}{\partial t}-\frac{u_z}{L_{gc}}+ 
\frac{\partial u_z}{\partial z} = 0\\
\frac{\partial u_z}{\partial t} = -(\eta-\zeta)\frac{u^2_{Ti}}{L_{gc}}
-u^2_{Ti}\frac{\partial\zeta}{\partial z}
-\frac{e}{M}\frac{\partial \phi}{\partial z}\\
\frac{\partial\zeta}{\partial t}-\frac{u_z}{L_{gc}} = -\gamma
\frac{\partial u_z}{\partial z}\\
\frac{\partial^2\phi}{\partial z^2}=\frac{e\overline{n}}{\varepsilon_0}
\left(e\phi/ T_e-\eta\right).
\label{full-set}
   \end{eqnarray}
We have $\gamma = 5/3$ for adiabatic ion dynamics. Alternatively, $\gamma = 1$ for isothermal dynamics and we have $\zeta=\eta$ there.
Taking a plane test-wave $\exp\big(-i(\omega t- k z)\big)$ we find a
dispersion  relation in the form
\begin{equation}
\omega = \frac{\sqrt{k} \sqrt{k {L_{gc}}+i } 
\sqrt{{C_s}^2+\gamma {u_{Ti}}^2 \left(k^2
   {\lambda_{De}}^2  +1\right)}}{\sqrt{i L_{gc} \left( k^2 {\lambda_{De}}^2
    +1 \right)}}.
\end{equation}
The result is local in the sense that we take 
${e^2\overline{n}}/{\varepsilon_0  T_e}$ fixed.

Assume the ratio of the Debye length and the vertical length scale $\lambda_{De}/L_{gc} \sim \epsilon^2$, where $\epsilon$ is a small dimensionless expansion parameter. We now expand the dispersion 
relation in powers of $\epsilon$. To lowest order we get the 
non-dispersive sound relation
$$
\omega \approx  k C_s,
$$
where the sound speed $C_s = \sqrt{(T_e+\gamma T_i)/M}\approx \sqrt{T_e/M}$ when $T_e \gg T_i$ as in our case.
To the next order in $\epsilon$ we find the additional term
$$
\frac{(T_e/M)(1+ik^3\lambda_{De}^2L_{gc})+\gamma u_{Ti}^2}{2L_{gc}C_s} = \frac{C_s}{2L_{gc}}+i\frac{k^3}{2}{C_s\lambda_{De}^2},
$$
see also Fig.~\ref{fig:lindisp}. We will use $C_s\approx \sqrt{T_e/M}$ in the following analysis. The linear 
differential equation for one of the plasma variables, say $u_z(z,t)$,
is obtained by the replacements $\omega\rightarrow i\partial/\partial t$ and
$k\rightarrow -i\partial/\partial z$.

\section{The Korteweg - de Vries equation}

\begin{figure}
\includegraphics[width=0.6\columnwidth]{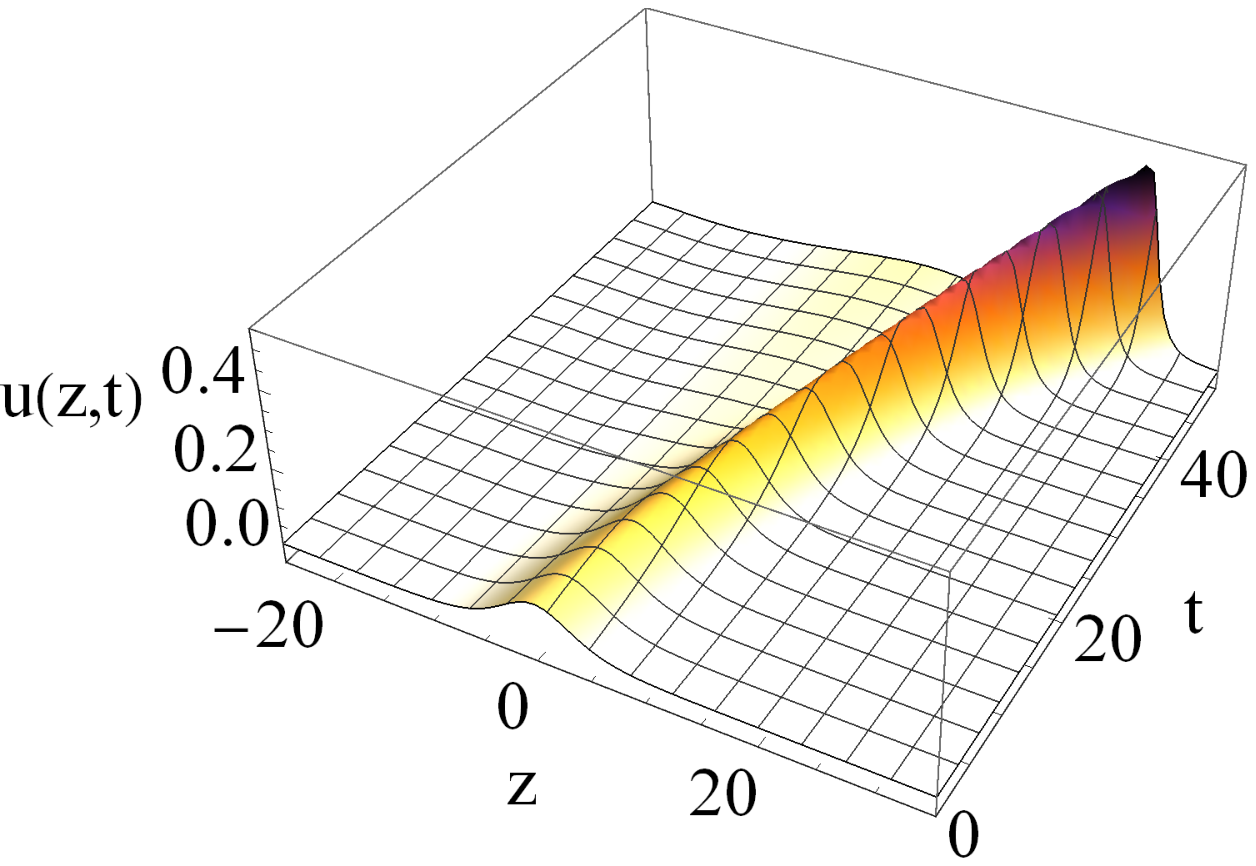}\\
\vspace{5mm}
\includegraphics[width=0.5\columnwidth]{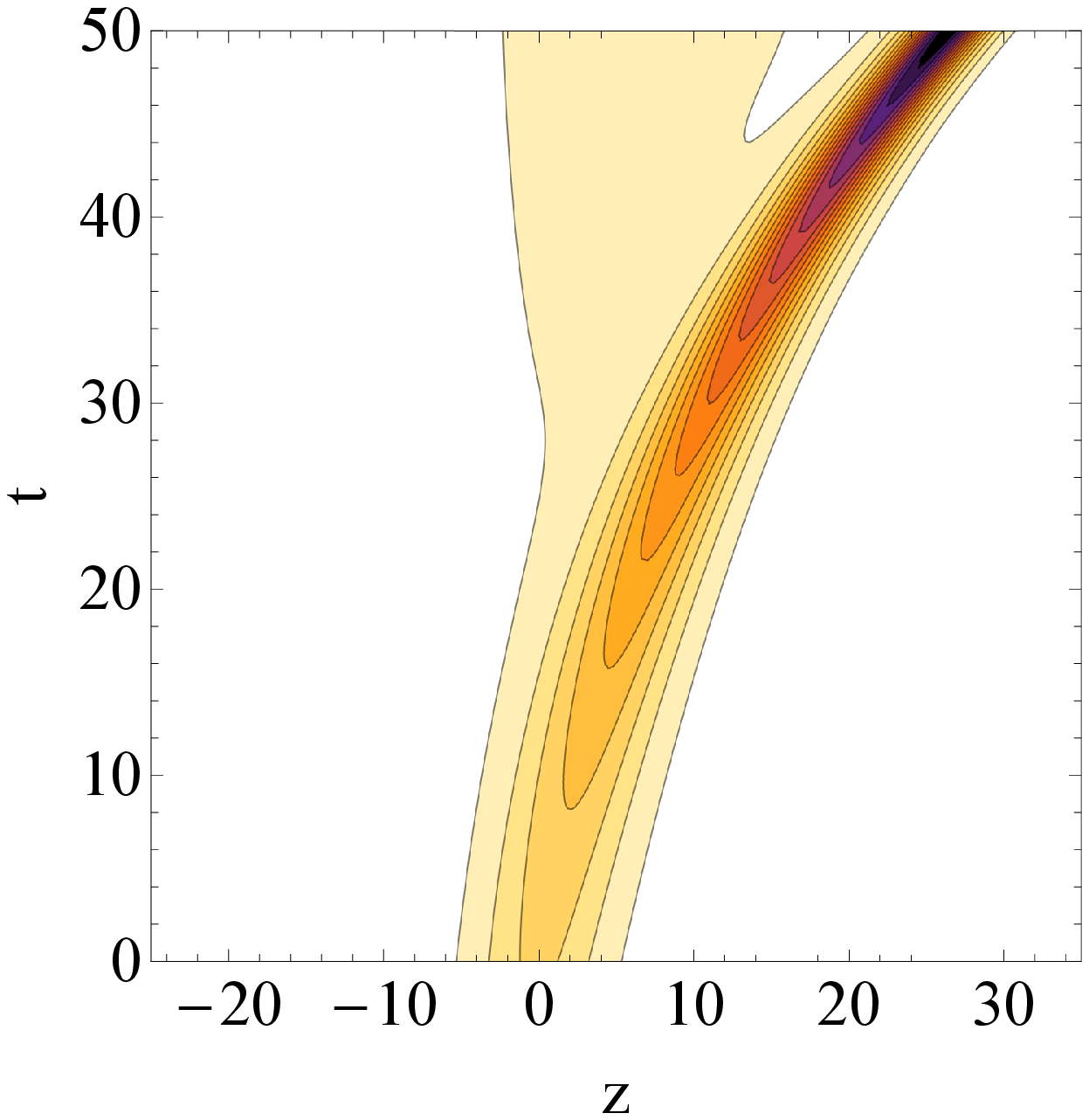}
\caption{\it Numerical solution of the modified 
KdV-equation (\ref{KdV_eq}) in the 
normalized form $\partial_tu+6u\partial_xu+\partial_{xxx}u = 
\gamma u$ with 
the initial pulse-shape being a soliton like (\ref{KdV_soliton}) with amplitude
$A_0 = 0.1$ and taking $\gamma = 0.025$.
The pulse is ``speeding up'' 
and becomes narrower as its amplitude increases
due to the growth term on the right hand side of (\ref{KdV_eq}). 
Note the formation of a ``plateau'' trailing the soliton.
There is an analytical basis also for this 
result \cite{karpman_1979,pecseli_2012}.
The figure refers to a frame of reference moving with the sound velocity.}  \label{fig:numsol}
\end{figure}

By a standard reductive perturbation analysis we can obtain a 
modified Korteweg - de Vries equation. 
Details of the method can be found
in a monograph \cite{nayfeh_1973}, and in 
particular also in the special issue on
``Reductive Perturbation Method for Nonlinear Wave Propagation'', 
Supplement of the Progress in Theoretical Physics, (1974) Vol.\ 55, 
published by the Research Institute for Fundamental Physics and the 
Physical Society of Japan. In the present analysis we
retain the lowest order correction in the dispersion relation
originating from Poisson 
equation, i.e.\ deviations from quasi-neutrality. We assume that ratio of the Debye length and the vertical length scale $\lambda_{De}/L_{gc}\sim\epsilon^2$ is 
of the same order as the fluid steepening nonlinearity in the expansion parameter. 
To lowest order in the small expansion parameter we therefore recover
the linear sound waves propagating in homogeneous plasmas. 
To next order we here
have dispersion, nonlinearity and the effects of density gradient entering
at the same level. We find a modified  KdV equation in the form
   \begin{equation}
\frac{\partial u_z}{\partial t}+(C_s+u_z)\frac{\partial u_z}{\partial z}+
\frac{1}{2}C_s\lambda_{De}^2\frac{\partial^3 u_z}{\partial z^3}= \frac{g}{2C_s}u_z.
\label{KdV_eq}
   \end{equation}
To lowest order (\ref{KdV_eq}) reproduces (\ref{simp_eq}) in the limit
of large $k_z$, i.e.\ for structures that are narrow in comparison with $L_{gc}$. 
The term on the right hand
side gives rise to a growth of the velocity perturbation associated
with a soliton or any other initial condition. 
The equation is here expressed for  the 
space-time varying velocity $u_z$, but  to lowest order we can use the relation
$e\phi/T_e\approx u_z/C_s$ to establish an equation for the 
electrostatic potential $\phi$.
Often the KdV-equation is written in the frame moving with the sound velocity.
Illustrative numerical solutions of (\ref{KdV_eq}) in this frame are shown in Fig.~\ref{fig:numsol}. This solution refers to the idealized case with the initial condition being an exact soliton solution which is usually considered in a perturbation analysis. In the absence of a density gradient it will propagate without deformation through the system. Note the formation of a plateau trailing the soliton for the inhomogeneous KdV equation. 
Ultimately also this plateau will break-up into a new small amplitude soliton as seen for large times in Fig.~\ref{fig:numsol}.

The KdV-equation is an approximation to the set of dynamic equations,
and the perturbation term  on the right hand side of (\ref{KdV_eq})
also represents an approximation to the full modification induced
by the plasma density gradient. We cannot expect an exact energy
conservation by (\ref{KdV_eq}).

The interest in these growing pulse solution is due to the possibility for soliton interactions
with plasma particles, in particular acceleration of particles 
by a first order Fermi
acceleration \cite{fermi_1949}. If applied to ionospheric conditions, such types of 
wave particle interactions can contribute to polar wind accelerations.

As well known, a KdV type equations describes unidirectional propagation of pulses.
We can formulate a slightly more general Boussinesq equation as shown in the Appendix.
This equation can have interest in its own right, but will not be used here.

\section{The homogeneous KdV equation}

For later use we first summarize some relevant results for KdV-solitons. The homogeneous KdV-equation in the general form
   \begin{equation}
\frac{\partial}{\partial t} u+\beta u\frac{\partial}{\partial z} u +
\alpha \frac{\partial^3}{\partial z^3} u =0 \, ,
\label{bare_KdV_eq}
   \end{equation}
has soliton solutions 
   \begin{equation}
u = A \, \mbox{sech}^2\left((z-U_st)\sqrt{A\, \beta/12\alpha}\right) \, ,
\label{KdV_soliton}
   \end{equation}
where the soliton velocity scales linearly with amplitude as $U_s=A\beta/3$.
The soliton width $\Delta = \sqrt{12\alpha/A\beta}$ 
scales inversely with the square root of the 
soliton amplitude. Large amplitude solitons are fast and narrow.
By the inverse scattering transform \cite{whitham_1974}
it can be demonstrated that any compact
initial perturbation will in time develop into one or more solitons 
followed by a low level of oscillations well described by the
linearized version of the KdV equation.

For the present analysis it is implicitly assumed that the soliton is
local in the sense that its width is smaller than the characteristic length
scale $\Delta\ll L_{gc}$. The parameters here are
$\alpha=\frac{1}{2}C_s\lambda_{De}^2$ 
and $\beta=1$ by  (\ref{KdV_eq}). As an estimate we have 
the velocity amplitude related to the density perturbation as
$A\approx C_s\delta n/n_0$. The requirement $\Delta\ll L_{gc}$ then imposes the 
restriction $\sqrt{6\lambda_{De}^2n_0/\delta n}\ll 2T/Mg$ or
$\delta n/n_0\gg\frac{3}{2}\lambda_{De}^2 M^2g^2/T^2$, 
which can be reduced to the simpler expression
$\delta n/n_0\gg\frac{3}{2}(\lambda_{De}/L_{gc})^2$. This requirement has 
to be imposed on the excitation of the soliton and the results 
are valid as long as the inequality is fulfilled, where $n_0$ then
refers to the plasma density at the soliton position.

A number of conservation laws are associated with the 
homogeneous KdV equation. A few  examples are \cite{drazin_johnson_1989}
\begin{eqnarray}
I_1&\equiv&  \int_{-\infty}^\infty \!u(z,t) dz \, ,\label{I_1}\\
I_2&\equiv& \int_{-\infty}^\infty \!\frac{1}{2}u^2(z,t) dz \, , \label{I_2}\\
I_3&\equiv&  \int_{-\infty}^\infty \! \left(\frac{\alpha}{3}u^3(z,t)+ 
\frac{1}{2}\left(\frac{\partial}{\partial z}u(z,t)\right)^2 \right)dz 
\, , \label{I_3}
\end{eqnarray}
where $I_2$ in particular is often associated with the energy of 
a perturbation. 
We note  here that this interpretation assumes homogeneous media.
For the soliton solution (\ref{KdV_soliton}) we find
$I_1 = 4 \sqrt{3 A \alpha /\beta}$ and $I_2= 4 A\sqrt{A\alpha/(3\beta)}$.
With an average position being $\int_{-\infty}^\infty z u(z,t) dz $ we
find a pulse velocity to be $\int_{-\infty}^\infty dz 
\,z \partial u(z,t)/\partial t$. 
For a soliton solution we readily find the velocity to be $U_s$
as given before.
The conservation laws (\ref{I_1})-(\ref{I_3}) are valuable for a subsequent perturbation analysis.

\section{Soliton perturbation analysis}

Korteweg-de Vries equations with perturbations have been studied 
in detail 
\cite{watanabe_1978,karpman_maslov_1977,karpman_1979,karpman_et_al_1980,wadati_akutsu_1984}. 
The simplest analysis
is based on conservation laws  \cite{watanabe_1978} and we
follow these.
Retaining the perturbation term on the right hand side of (\ref{KdV_eq})
the conservation laws become 
\begin{eqnarray}
\frac{d I_1}{dt}&=& \frac{g}{2C_s}I_1,\label{I_1a},\\ 
\frac{d I_2}{dt}&=& \frac{g}{C_s}I_2, \label{I_2a}
\end{eqnarray}
giving $I_1(t)=I_1(0)\exp(\frac{1}{2}t g/C_s)$ and 
$I_2(t)=I_2(0)\exp(t g/C_s)$. Taking the initial perturbation
to have a soliton shape we have $I_1(0)=4 \sqrt{3 A_0 \alpha /\beta}$ 
and $I_2(0)= 2 A_0\sqrt{A_0\alpha/(3\beta)}$. 

Starting the problem with a 
soliton solution we assume that it 
at all times retains its soliton shape:  for slow variations
this assumption is justified by the 
inverse scattering transform. Since the soliton
is a one parameter solution we expect that we at all times can
quantify its characteristics by its amplitude. Velocity and width
follows from this amplitude. A small non-soliton part, $u_{ns}$ 
as seen developing in
Fig.~\ref{fig:numsol}, is necessary to accommodate the difference 
between the entire solution $u_z(z,t)$ and the time evolving soliton
part $u_s$.
Since the non-soliton part has a small amplitude it has a small velocity
in the frame of reference moving with $C_s$
and it will be a ``tail'' following the soliton: we assume that the 
overlap between these two components of  $u_z(z,t)$ is negligible
implying $u_s (z,t)u_{ns}(z,t)\approx 0$. The plateau starts at $z\approx 0$
in the moving frame and ends at the soliton position 
in the moving frame $\langle z(t)\rangle=\int_0^tU_s(\tau)d\tau$
in terms of the soliton velocity $U_s(t)=A(t)\beta/3$. We let the plateau be 
characterized by a spatially
averaged amplitude $\xi(t)$, so that $I_1(t)\approx 
\langle z(t)\rangle \xi(t)+ 4 \sqrt{3 A(t) \alpha /\beta}$
and $I_2(t)\approx \langle z(t)\rangle \xi^2(t) 
+2 A(t)\sqrt{A(t)\alpha/(3\beta)}$. Together with the first two conservation
laws we have two equations for the two unknowns, $A(t)$ and $\xi(t)$,
since the time varying soliton velocity and thereby $\langle z(t)\rangle$ are
determined through the soliton amplitude $A(t)$. Assuming
$\xi$ to represent a small correction, we ignore terms containing $\xi^2$.
From the expression for $I_2(t)$ we then have 
$$
A(t)\approx A(0)\exp\left(t \frac{2g}{3C_s}\right).
$$
As the length of the plateau increases, it can itself break up into solitons. As a consequence a local density and thereby also a local potential minimum develops behind the soliton which can subsequently participate in the kinetic particle interactions.

The soliton position in the moving frame is found by
\begin{eqnarray*}
\langle z(t)\rangle&\approx& \int_0^tU(\tau)d\tau=
\frac{A(0)\beta}{3}
\int_0^t\exp\left(\tau \frac{2g}{3C_s}\right)d\tau\\
&=& \frac{A(0)C_s\beta}{2g}
\left(\exp\left(t \frac{2g}{3C_s}\right)-1\right).
\end{eqnarray*}
To transform to the fixed frame we have to add $t C_s$.

Using the results for $I_2(t)$ we can obtain an approximate expression for
the kinetic energy of the system as
\begin{equation}
{\cal E}_k\approx M
n_0\exp\left(-\frac{t g}{C_s}\right)I_2(t)=\mbox{constant}
\label{kin_energy}
\end{equation}
at any time $t$, recalling that this expression is
meaningful only in the rest frame. We approximated the
soliton position as $z\approx t C_s$ in $\overline{n}(z) = n_0\exp\left(-zg/C_s^2\right)$. For large times we find
${\cal E}_k\rightarrow$~constant to the lowest approximation
as long as  $t C_s\gg\langle z(t)\rangle$.
The contribution of the electrostatic field to the total energy can be
determined the same way.

Given $A = A(t)$ we can determine the average amplitude of the
non-soliton part $\xi(t)$ by the expression for $I_1(t)$.
After some algebra we find
\begin{eqnarray*}
\xi(t)&=&\frac{8g\sqrt{3\alpha}}{C_s\beta\sqrt{A(0)\beta}}
\frac{\exp(\frac{tg}{2C_s})-\exp(\frac{tg}{3C_s})}{\exp(\frac{t2g}{3C_s})-1}\\
\lim_{t\rightarrow\infty}\xi(t)&=&\frac{8g\sqrt{3\alpha}}{C_s\beta\sqrt{A(0)\beta}}
\exp\left(-\frac{tg}{6C_s}\right).
\end{eqnarray*}
At large times the soliton amplitude is exponentially large and so 
is its velocity. Asymptotically, the non-soliton tail is stretched 
out to have a small amplitude. A large initial amplitude $A(0)$ has the same
effect.

For the entire energy budget we have to include both the 
soliton and the non-soliton parts. For interaction with particles, we need to
be concerned only with the soliton part since it has the dominant amplitude.

\section{Interaction between solitons and ions}

The foregoing analysis emphasizes fluid models. The problem of
plasma wave propagation in gravitational field in a horizontally striated
plasma environment has previously \cite{parkinson_schindler_1969,liu_1970}
been studied by linear kinetic models, including effects of Landau damping.
The time interval where linear Landau damping is however of minor relevance
for the problem when the nonlinear soliton evolution is considered.
To see this we introduce a few relevant time-scales: 1) a linear
pulse time-scale $\tau_L = \Delta/C_s$, which corresponds to the linear
sound dispersion relation. 2) we have a nonlinear soliton time scale $\tau_S$
which accounts for the time it takes a soliton to move its own width
due to the nonlinear velocity correction \cite{lynov_et_al_1979_report,karpman_et_al_1980}, 
i.e., the motion in the
frame moving with the sound speed $C_s$, giving $\tau_S = \Delta/U_s$
where $\tau_S\gg\tau_L$. In classifying the interaction between
particles and wave-pulses we have a time of linear or resonant
interaction $\tau_R = \Delta/\sqrt{2e\Psi/M} \sim \tau_S$ 
where $\Psi$ is the peak value
of the electrostatic potential for the soliton. The velocity
interval for resonant wave-particle interaction is
$\left[C_s + U_s - \sqrt{2e\Psi/M};C_s + U_s + \sqrt{2e\Psi/M}\right]$
specifying the role of the soliton amplitude. 
The linear Landau
damping is associated with transiting particles \cite{chen_1984}.

We thus distinguish two parameter ranges. 1) 
Times $t < \tau_R$ where linear Landau
damping dominates and soliton dynamics is of minor importance. 2)
Times $t > \tau_R\sim\tau_S$ where soliton dynamics is important and the
interaction between the nonlinear sound pulse and particles
is (in our case) dominated by reflected ions.

To describe the propagation of weakly nonlinear sound waves in a kinetic model, several authors
\cite{ott_sudan_1969,vandam_taniuti_1973,saito_nakamura_2003,sikdar_khan_2017} have proposed a modified KdV-equation in the form 
\begin{equation}
\frac{\partial}{\partial t} \phi+\beta \phi\frac{\partial}{\partial z} \phi +
\alpha \frac{\partial^3}{\partial z^3} \phi + \frac{s}{\pi}
{\cal P}
\int_{-\infty}^\infty \frac{1}{z-z^\prime} 
\frac{\partial\phi}{\partial z^\prime}  dz^\prime = 0  ,
\label{linlandau_KdV_eq}
\end{equation}
with ${\cal P}$ denoting the principal value of the integral and $\beta$, $\alpha$, and $s$ being suitably defined constants. 
The nonlocal integral term accounts for the linear Landau damping
here and in a number of related studies 
\cite{ichikawa_taniuti_1973,dysthe_pecseli_1977},
and the equation is thus valid for the time-range 1) discussed before.
In this time interval the solitons properties had little time to be manifested
in any significant manner. The applicability of 
(\ref{linlandau_KdV_eq}) is limited as far as the nonlinear
soliton dynamics are concerned, although the equation 
had received attention in the past.

Many of the foregoing results had applications for general KdV-equations.
The present problem concerns acceleration of plasma particles  
by solitons propagating in 
gravitational plasmas with a vertical density gradient. For this
case we have $\beta=1$ in (\ref{bare_KdV_eq}) while
$\alpha= \frac{1}{2}C_s\lambda_{De}^2$, see also (\ref{KdV_eq}).
The simple model used here assumes electrons to be an isothermally 
Boltzmann distributed fluid at all time, with electron
inertia effects ignored. The only plasma particles we need 
to be concerned with are the ions.



Given a soliton with velocity amplitude
$A(t)$ we have the corresponding peak potential amplitude to be $\Psi(t)=
A(t)( T_e/e)/C_s$. The velocity interval for resonant ion interaction has then the form 
$\left[ C_s(1+ \frac{1}{3}e\Psi(t)/ T_e)-U_R;
C_s(1+ \frac{1}{3}e\Psi(t)/ T_e)+U_R\right]$. 
Particles slower than $C_s(1+\frac{1}{3}e\Psi(t)/ T_e)$ 
give up energy, while faster 
particles receive energy from the moving soliton.
For the ions overtaking the soliton there would be a 
slight correction due to the plateau, but this will be ignored here.
We here introduced $C_s(1+\frac{1}{3}e\Psi(t)/
 T_e)$ for the rest frame soliton velocity
so that $U_s = \frac{1}{3}eC_s\Psi(t)/
 T_e$. We find that $\tau_S/\tau_R\sim C_s/\sqrt{2e\Psi/M}\gg 1$.
When the soliton dynamics is important, the linear Landau damping is of
minor concern. The important soliton-particle interaction
is caused by reflected particles, which is a nonlinear effect.

The following discussion will be based on energy conservation between
a system consisting on a soliton and plasma particles. 
We will use the capital letter $U$ denoting the
$z$-component of one ion as distinguished from a fluid velocity. The 
kinetic + electric energy
of an ion acoustic soliton
in a gravitational field is \cite{pecseli_2012}
\begin{equation}
{\cal E}\approx 4\sqrt{\frac{2}{3}} \left(\frac{e\Psi(t)}{ T_e}\right)^{3/2}
{n_0 \exp\left(-\frac{t\, g}{C_s}\right)  T_e\lambda_{De}} \, ,
\label{soliton_energy}
\end{equation}
see also (\ref{kin_energy}). Upon interaction
with a soliton moving at velocity $U_s$, an ion changes its initial
velocity $U$ by the amount $2U_s$.
The energy gain by such an interacting (i.e.\ resonant) 
ion is $2 MU_s(U_s-U)$, assuming the interaction
to be perfectly elastic. A negative ion velocity (counter
propagating particles) gives net particle energy gain, positive ion
velocities (overtaking collisions) give energy loss. 
The flux of these interacting ions is at some
vertical position $z$ given as $|u-U_s|\,\overline{n}(z)f_0(u)$, 
where $f_0 (u)$ is the normalized background ion velocity
distribution function, 
$\int_{-\infty}^\infty f_0(u) du = 1$. Consequently at a time where the
soliton has arrived at a position $z = C_s t$, we can write the
energy gain by resonant ions per unit time as
\begin{equation}
\frac{d{\cal E}_{res}}{dt} = 2 M U_s n_0
\exp\left(-\frac{t\, g}{C_s}\right) \int_{U_{min}}^{U_{max}}
(U_s-U)\, |U-U_s|\, f_0(U)dU.
\label{part_energyloss}
\end{equation}
The  integration limits are $(U_{min};U_{max}) = 
\left(U_s -\sqrt{2e\Psi(t)/M};U_s +\sqrt{2e\Psi(t)/M}\right)$. 

We now equate
this change in energy per time-unit with the negative time derivative of the change 
in soliton energy obtained from (\ref{soliton_energy}). 
The foregoing arguments assume that the soliton amplitude $\Psi(t)$ changes only little during the transit time of an ion.

The foregoing analysis refers to one soliton interacting with particles. For larger soliton densities, 
solitons can 
interact due to mutually reflected particles \cite{honzawa_1982}.
A statistical analysis of such many-soliton cases has also been
suggested \cite{dysthe_pecseli_trulsen_1986}.

\begin{figure}[htb]
\begin{center}
\includegraphics[width=0.45\textwidth]{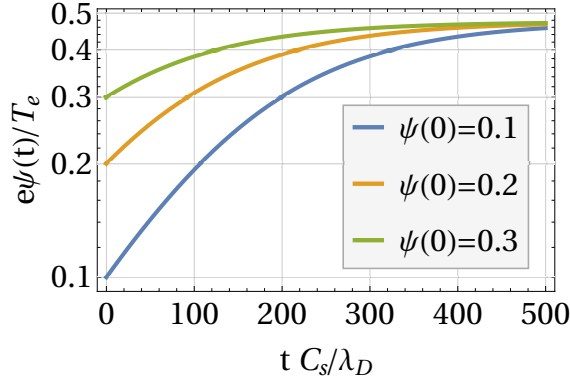}
\caption{\em Illustration 
of the normalized amplitude variation of an ion acoustic soliton
as described by (\ref{fin_solprt_int}) for three different initial soliton amplitudes, $\Psi(0) = 0.1, 0.2,$ and $0.3$.
The figure uses normalized units, with a logarithmic vertical 
axis and $C_s \equiv \sqrt{ T_e/M}$. We have here
$T_e/T_i = 10$ and a dimensionless
``gravity parameter'' $g\lambda_{De}/C_s^2 = 0.01$. 
Less interesting solutions with larger initial amplitude, $\Psi(0) > 0.5$ for the present parameters,
damp out to reach the same asymptotic level as
shown in the figure for the other 
amplitudes.}\label{fig:soliton_damp}    
\end{center}
\end{figure}

We have found the energy gained or lost
by ions accelerated or decelerated by a soliton. 
By energy conservation we know that this energy is lost from the 
soliton. All soliton parameters can be expressed by the maximum
soliton amplitude $\Psi(t)$
for the KdV-soliton discussed here.
Since a relation between the soliton parameter and 
the soliton energy is known we can obtain an equation 
for $\Psi(t)$. The rate of change of soliton energy for varying 
$\Psi(t)$ is
\begin{equation}
\frac{d{\cal E}}{dt} = 4\sqrt{\frac{2}{3}}\sqrt{\frac{e\Psi(t)}{T_e}}\,
n_0 \exp\left(-\frac{t\, g}{C_s}\right) T_e\lambda_{De} 
\left(\frac{e}{ T_e}
\frac{d\Psi}{dt} - \frac{g}{C_s}  \left(\frac{e\Psi(t)}{T_e}\right)
\right).
\label{solprt_int}
\end{equation}

Equating (\ref{solprt_int}) and (\ref{part_energyloss}) we note
that the exponential factors cancel and obtain after some algebra
\begin{equation}
\frac{d}{dt}\frac{e\Psi(t)}{T_e} =
\frac{g}{C_s} \frac{e\Psi(t)}{T_e} + \frac{1}{3}\sqrt{\frac{3}{2}}
\sqrt{\frac{T_e}{e\Psi(t)}}\frac{MU_s}{T_e\lambda_{De}}
G\big(U_s,\Psi(t)\big) \, ,
\label{fin_solprt_int}
\end{equation}
with
$$
G(U_s,\Psi(t)) = \int_{U_s}^{U_s+\sqrt{2e\Psi(t)/M}}(u-U_s)^2 f_0(u) du
+ \int_{U_s}^{U_s-\sqrt{2e\Psi(t)/M}}(u-U_s)^2 f_0(u) du \, ,
$$
recalling here that $U_s$ depends also on $\Psi(t)$, in general. 
For Maxwellian
distributions, we can express $G(U_s,\Psi(t))$ in terms of error
functions. A numerical solution of (\ref{fin_solprt_int})
is shown in Fig.~\ref{fig:soliton_damp} assuming a 
Maxwellian distribution for $f_0(u)$.
We find that a soliton with small initial amplitude 
has its peak potential amplitude
increasing according to the ``fictitious growth'', but at some time
its amplitude is sufficiently large to have it interacting significantly
with the ions. The growth is then arrested, 
eventually to reach a saturated level.
The saturation level and the time evolution in general depends 
on the electron-ion temperature ratio $T_e/T_i$ as well as $g/C_s$.
If $T_e/T_i$ is reduced, the ion sound speed becomes closer to 
the ion thermal velocity and the soliton-particle interaction 
becomes stronger giving a lower saturation level. The asymptotic
saturation level for the peak soliton potential does not in general
have any simple analytical expression. For the net soliton energy
we have ${\cal E}(t\rightarrow\infty)\rightarrow 0$ when the 
soliton-particle interaction is taken into account for a stable plasma,
e.g.\ a Maxwellian. The net kinetic energy gained by the particles
equals the initial soliton energy. The density gradient acts 
as a ``catalyst'' mediating the energy transfer.

\subsection{Analytical approximations}

In order to obtain some quantitative results, we make a series expansion of
$G(U_s,\Psi(t))$ in (\ref{fin_solprt_int}), 
where we here let the soliton velocity be a constant
$U_s\approx C_s$ since the correction varies only 
with ${\Psi(t)}$ which was assumed to be small anyhow.
We then have 
\begin{eqnarray*}
G(U_s,\Psi(t)) &=&  2\left(\frac{e\Psi(t)}{M}\right)^{2}f_0^{(1)}(C_s)\\
&&\times\left(1+4\sum_{n=3}^\infty\frac{(2n-1)(2n-2)}{(2n)!}
\frac{f_0^{(2n-3)}(U_s)}{f_0^{(1)}(U_s)}
\left(\frac{2e\Psi(t)}{M}\right)^{n-2}\right)  ,
\end{eqnarray*}
or
\begin{eqnarray*}
G(U_s,\Psi(t))&=&2\left(\frac{e\Psi(t)}{ T_e}\right)^{2}C_s^4f_0^{(1)}(C_s)\\
&&\times\left(1+4\sum_{n=3}^\infty
\frac{(2n-1)(2n-2)}{(2n)!}
\frac{f_0^{(2n-3)}(U_s)}{f_0^{(1)}(U_s)}
\left(\frac{2e\Psi(t)}{ T_e}\right)^{n-2}C_s^{2n-2}\right) ,
\end{eqnarray*}
where $f_0^{(m)}$ denotes the $m$-th derivative of $f_0(u)$. To lowest order,
we can write the relation (\ref{fin_solprt_int}) as
$$
\frac{d}{dt}\frac{e\Psi(t)}{ T_e}=\frac{g}{C_s} \frac{e\Psi(t)}{ T_e}+
\sqrt{\frac{2}{3}}
\frac{C_s^3}{\lambda_{De}}
\left(\frac{e\Psi(t)}{ T_e}\right)^{3/2}f_0^{(1)}(C_s) \, ,
$$
which can  be integrated to give
\begin{equation}
\frac{e\Psi(t)}{T_e} = 
\frac{(g/C_s)^2{e\Psi(0)}/{T_e}} {\left(\left({g}/{C_s} - \nu\sqrt{{e\Psi(0)}/
{T_e}}\right)\exp\left(-\frac{1}{2} t g/C_s\right) + \nu\sqrt{{e\Psi(0)}/{T_e}}\right)^2},
\label{part_reflect_damp}
\end{equation}
where the damping constant is 
$$
\nu = -\sqrt{\frac{2}{3}}
\frac{C_s^3}{\lambda_{De}}f_0^{(1)}(C_s)  .
$$
When $f_0(u)$ is a Maxwellian, for instance, we have $f_0^{(1)}(C_s) < 0$
giving $\nu > 0$,
and the soliton amplitude reaches an asymptotic level \cite{lynov_et_al_1979_report}. 
When $\nu > 0$, the model (\ref{part_reflect_damp}) gives the
asymptotic saturation level for the soliton amplitude
as $e\Psi(\infty)/ T_e = (g/C_s)^2/\nu^2$, independent of the initial
value $\Psi(0)$.

For a linearly unstable plasma where $f_0^{(1)}(C_s) > 0$, giving $\nu < 0$,
we can find an ``explosive'' condition by  (\ref{part_reflect_damp})
where $\Psi(t)$ can be diverging within a finite time $\tau_c$
given implicitly by $\left({g}/{C_s}-\nu\sqrt{{e\Psi(0)}/
{ T_e}}\right)\exp\left(-\frac{1}{2}\tau_c \,g/C_s\right)=
-\nu\sqrt{{e\Psi(0)}/{ T_e}}$. Such a ``bump-on-tail'' 
condition for the net ion velocity distribution
can, for instance, be realized by an accelerated lighter ion component 
constituting the polar wind mentioned before.

Unfortunately, the compact result (\ref{part_reflect_damp}) has 
limited applicability \cite{lynov_et_al_1979_report}. 
This limitation can be illustrated by considering
the next correction term in the series expansion in $G(U_s,\Psi(t))$. In this case we have
\begin{equation}
G(U_s,\Psi(t))\approx 2\left(\frac{e\Psi(t)}{M}\right)^{2}f_0^{(1)}(U_s) 
\left(1+\frac{1}{9}\frac{f_0^{(3)}(U_s)}{f_0^{(1)}(U_s)}
\frac{2e\Psi(t)}{M}\right).
\label{correction-term}
\end{equation}

For an order of magnitude estimate we can use a Maxwellian ion velocity distribution, $f_0(u) = (2\pi\sigma)^{-1/2}\exp (-u^2/2\sigma)$, with
$\sigma\equiv   T_i/M \ll C_s^2$. For the last correction term in the
parenthesis 
to be small we require $(C_s^2/\sigma)(e\phi_0/ T_i)\ll 5$, which is only marginally 
realistic in natural conditions, when we at the same time
require that the nonlinearities should be manifested in a reasonable time,
i.e., that the soliton time should be moderate. It is most likely that 
(\ref{fin_solprt_int}) has to be solved numerically for realistic and relevant cases as in  Fig.~\ref{fig:soliton_damp}. We find that the saturation level $e\Psi(\infty)/ T_e$ found by (\ref{part_reflect_damp}) to be an overestimate, in general.

\begin{figure}[htb]
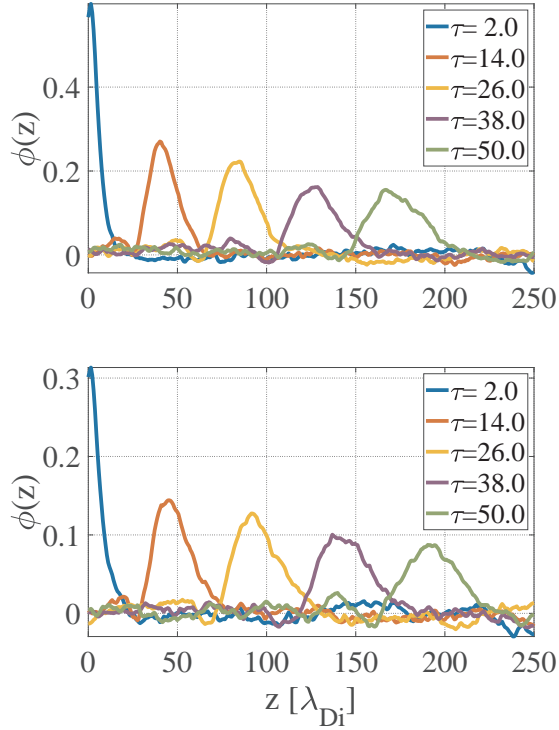

\begin{center}
\includegraphics[width=0.45\textwidth]{{{gravity_pulse2d_a0.25G0Th11L500s-fig1}}}\\
\includegraphics[width=0.45\textwidth]{{{gravity_pulse2d_a0.1G0Th16L1000s-fig1}}}
\caption{\em Spatial variations of propagating solitons taken at selected time 
steps for the reference case with no gravitational field, 
$G = 0$. We have $T_e/T_i = 10$ in the top and $T_e/T_i = 15$
in the bottom figure, respectively. The damping is due to ion 
Landau damping, which is strongly reduced by the
increased temperature ratio in the second case. The 
externally imposed excitation
amplitudes are $0.25$ and $0.1$ for the two cases.
The first narrow pulse on the figure is a part of the initial
excitation. The difference in propagation velocity is due to the change in the sound speed.}\label{fig:numsim-ref}    \end{center}
\end{figure}

\begin{figure}[htb]
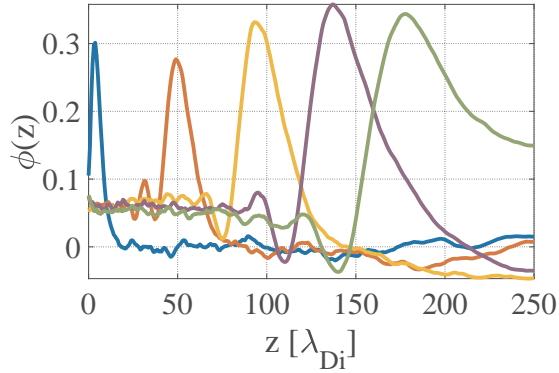

\begin{center}
\includegraphics[width=0.45\textwidth]{{{gravity_pulse2d_a0.05G0.5Th16L1000s-fig1}}}
\caption{\em Spatial variations of propagating solitons taken at selected time 
steps $\tau = 14, 26, 38$, and $50 \,\Omega_{pi}$, 
with $G = 0.5$ in normalized units.
We have $T_e/T_i = 15$. Comparing with Fig.~\ref{fig:numsim-ref}  
we note an initially increasing amplitude due the fictitious growth
induced by the plasma density gradient in the gravitational field.
The first narrow pulse at $\tau = 2\,\Omega_{pi}$
on the figure is also here a part of the initial
excitation.}\label{fig:numsim-1}    
\end{center}
\end{figure}

\section{Numerical simulation results}

Our hybrid code with kinetic ions and mass-less isothermally Boltzmann 
distributed electrons assumes 
$n_e = n_{0}\exp\left({e\phi}/{ T_e}\right)$ from the outset,
implying that Poisson's equation becomes nonlinear in the
present problem. 
The ion component responds to the collective electric fields and to an imposed constant vertical gravitational field.
The numerical simulation results allow for deviations from
quasi neutrality since Poisson's equation
is explicitly included. 
The initial conditions can be chosen to have characteristic
scale lengths much larger than $\lambda_{De}$ so that quasi neutrality
can be assumed, but at later times we can find smaller scales to develop
and deviations from quasi neutrality can become important. In this
limit (\ref{poisson}) will be relevant, and the expression is 
implemented in our Particle in Cell (PIC) code.  Details of the code are described elsewhere \cite{guio_et_al_2003,guio_pecseli_2016}.
Most studies of KdV-solitons are based on models in
strictly one spatial dimension. To make the analysis somewhat more 
physically relevant we consider a two dimensional  magnetized system.
A generalization to a fully 3 dimensional system will in
our case not bring any new features to the problem.
The basic plasma parameters are chosen to be consistent with the
assumptions of the model, i.e., $\Omega_{ci} > \Omega_{pi}$.
Assuming an enhanced electron temperature in a central magnetic flux tube
we can also here derive a KdV-equation for a lowest order radial eigenmode.
The present analysis is related to studies of weakly nonlinear electrostatic 
Trivelpiece-Gould modes in a magnetized plasma wave-guide
\cite{manheimer_1969}. Details
of the analytical model  used here
are given elsewhere \cite{guio_pecseli_2016}. The basic analysis
gives an equation for ``simple waves'' \cite{blackstock_1972}, which is
subsequently generalized by introducing dispersion and the
effect of gravity to give a modified Korteweg-de Vries equation.

\begin{figure}[htb]
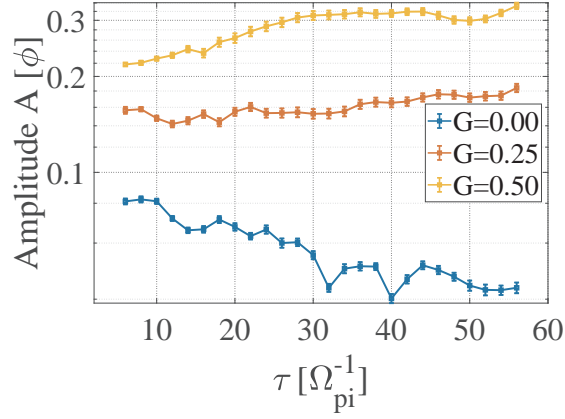

\begin{center}
\includegraphics[width=0.45\textwidth]{{{FitA0.05G0.00Th16L1000s-FitA0.05G0.25Th16L1000s-FitA0.05G0.50Th16L1000s-fig2}}}
\caption{\em Time evolution of the peak value of
the soliton potential amplitudes $A(\tau)$ in computational units shown on
a logarithmic scale. For the largest value of the gravitational acceleration $G = 0.5$, in computational units, we have an initial time interval with a
near exponential growth. The ultimate saturation is due to ions reflected by the large amplitude sound 
pulse. Also shown is the time evolutions for $G = 0.25$ and $G = 0$. We have $T_e = 15\, T_i.$}\label{fig:numsim-2}    
\end{center}
\end{figure}

\begin{figure}[htb]
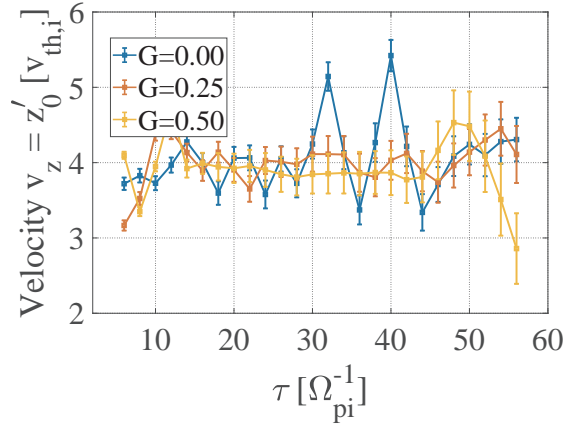

\begin{center}
\includegraphics[width=0.45\textwidth]{{{FitA0.05G0.00Th16L1000s-FitA0.05G0.25Th16L1000s-FitA0.05G0.50Th16L1000s-fig3}}}
\caption{\em  Time variation of the soliton velocity shown in units of the ion thermal velocity. The velocity is obtained by $z_0^\prime \equiv dz_0(\tau)/dt$, with $z_0(\tau)$ being the position of the soliton maximum.}\label{fig:numsim-3}    
\end{center}
\end{figure}

\begin{figure}[htb]
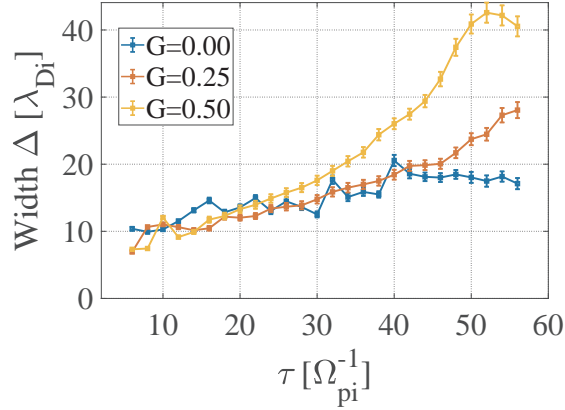

\begin{center}
\includegraphics[width=0.45\textwidth]{{{FitA0.05G0.00Th16L1000s-FitA0.05G0.25Th16L1000s-FitA0.05G0.50Th16L1000s-fig4}}}
\caption{\em Time variation of the soliton width, $\Delta(\tau)$, measured in units of the ion Debye length.}\label{fig:numsim-4}    
\end{center}
\end{figure}

\begin{figure}[htb]
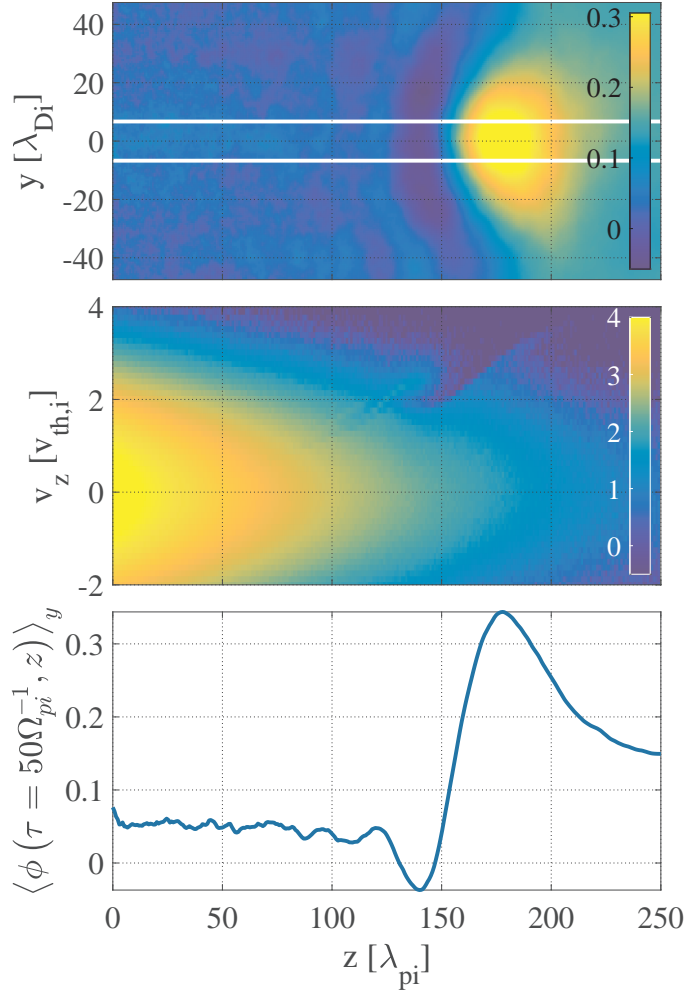

\begin{center}
\includegraphics[width=0.55\textwidth]{{{gravity_pulse2d_a0.05G0.5Th16L1000s-fig5}}}
\caption{\em Color coded spatial variation of the soliton variation in configuration space with linear color scale in a) while in b) we have the phase space variation of the same structure with a logarithmic color scale, 
here averaged over the central spatial region. In c) we have the corresponding
spatial potential variation also averaged over the central part of the plasma column. All figures refer to a selected time step, $\tau = 50\, \Omega_{pi}^{-1}$. The two white lines in the top figure indicate the central
``channel'' with the enhanced electron temperature. The gravitational acceleration points in the negative $z$-direction.}\label{fig:numsim-5}    
\end{center}
\end{figure}

Results from numerical simulations are shown in 
Figs.~\ref{fig:numsim-ref}-\ref{fig:numsim-5}. The figures show only the part $0 \leq z \leq 250\lambda_{Di}$ of a simulation
domain of $500 \lambda_{Di}$. In 
order to improve the signal-to-noise ratio 
in Fig.~\ref{fig:numsim-ref}, we averaged 4 results from
simulations with different initializations of the random number 
generators distributing the simulation particles.
In Fig.~\ref{fig:numsim-ref} we show two results, one reference case with
no gravity and a temperature ratio of $T_e/T_i = 10$, and a second
case with a constant gravitational acceleration $G = 0.5$ in our normalized
units and a temperature ratio of $T_e/T_i = 15$. In the first case
we observe the ion Landau damping, which is strongly reduced in the
second case due to the larger ion sound speed $C_s = \sqrt{(T_e + \gamma T_i)/M}$.
The solitons are shown at the same times, and the difference in their
basic velocity is noticeable. The nonlinear velocity correction is small in comparison.

The peak value of the soliton amplitude variations are shown in Fig.~\ref{fig:numsim-2}. We note in particular that 
this variation is exponential only for a restricted initial time interval, even for the case without
gravitational forces, $G = 0$. 

For the gravitational case, $G\neq 0$, we find an amplitude increase as predicted by
the simple model. Eventually the soliton amplitude reaches a level 
where it interacts strongly with the particles and find an amplitude saturation for large times. We note the formation
of a "fore-runner" or precursor in front of the
soliton for increasing times, see Fig.~\ref{fig:numsim-1} 
for instance. This is caused by the ions reflected 
and energized by the propagating soliton.

The soliton velocity as given in Fig.~\ref{fig:numsim-3}  is nearly constant, corresponding
to the ion sound speed for the given conditions. Some "spikes" for the case with $G = 0$ are due to inaccuracies in 
the numerical fitting procedures. The nonlinear velocity correction is small.

The variation of the soliton width is shown in Fig.~\ref{fig:numsim-4}. For the case with $G = 0$ we find that the
amplitude-width scaling predicted by the KdV-equation is qualitatively correct. When $G\neq 0$ we do not find this
agreement. Most likely this disagreement is caused by the uncertainty in defining a proper soliton width
when we have a precursor in the form of particles (in our case ions) reflected by the soliton.

The full configuration and phase space information is given in
Fig.~\ref{fig:numsim-5} for a late time $\tau= 50\, \Omega_{pi}$ in the evolution. The bulk plasma density increases when
moving from large $z$ towards $z = 0$ consistent with a balance between the gravitational and plasma pressure forces
as discussed in obtaining (\ref{iontemp-gravity}), for instance.

The localized density depletion forming behind the soliton gives a potential well that can trap particles to form a phase space vortex there. In Fig.~\ref{fig:numsim-5}  we find the formation of such a phase space vortex behind the solitary form. These vortex-like structures have been found experimentally first in electron phase space \cite{saeki_et_al_1979} and then also in ion phase space \cite{pecseli_trulsen_armstrong_1984}. See also a summary \cite{guio_et_al_2003}. 
In front of the soliton we note the population of reflected ions: visually, it appears similar to the "snow plow" effect found in front of shocks propagating in for instance coaxial plasma accelerators \cite{chang_1961,hart_1964}. The solitary pulse is
excited in the central part of the plasma (between the two white lines in the top figure). The boundary conditions for the electric field makes the pulse spread in the $y$-direction across magnetic field lines into the surrounding plasma
where $T_e = T_i$.

A number of observations can be made on the basis of the simulation results.
Some basic features predicted by the KdV equation are thus recovered,
i.e., we find a growth of pulse amplitude as it propagates in the 
direction opposite to the gravity direction. Fine details like
the amplitude-width soliton relation are however not recovered.
The soliton  amplitude-width relation is qualitatively satisfied only
for the case where we set gravitational acceleration $G=0$. For
this particular case, the soliton deformation is small, and it is
easier to make a soliton fit to the simulation curve. When we have
a significant amount of reflected particles and at the same time
formation of a trailing phase space vortex, it becomes difficult
to find a proper identification of the width of a pulse and
a local soliton property can no longer be demonstrated.

\section{Conclusion}

In the present study we analyzed weakly nonlinear ion acoustic sound pulses propagating
in a gravitational plasma with an isothermal equilibrium. For this 
inhomogeneous system we can solve the linearized wave propagation problem in a fluid model
analytically and find a ``fake'' instability leading to growth for
waves and pulses propagating in the anti-gravity direction. This is not a true instability \cite{dysthe_misra_trulsen_1975}
and has its origin in conservation of the flux
of wave energy density in a medium with varying density. 
The potential of the wave has an increasing amplitude at increasing altitudes
and becomes effective in reflecting particles. Ultimately, all wave
energy is transformed to particle energy. The gravitational field thus
serves as a ``catalyst'' in the transformation. We believe this to be a new observation. The system is energy conserving
and we can not gain particle energy exceeding what was
present in the electrostatic pulse at $z = 0$. Significant
particle acceleration is found only in cases where we
have large net energy in the injected pulses. If the ideas
outline in the present study are applied to the polar ionosphere with vertical or nearly vertical magnetic field lines, we anticipate that relevant conditions are found for unstable E-region conditions due to a two stream instability, for instance  \cite{kelley_1989}.

To give the problem an analytical basis we derived an approximate
model in terms of a modified Korteweg-de Vries equation. We studied
the propagation and deformation of soliton solutions for this equation.
Some basic features of the numerical results are explained by the 
model equation also concerning the energy exchange between solitons and plasma ions.
For the entire energy budget we have to include both the 
soliton and the non-soliton parts, such as plateau and tail. For interaction with particles, we need to
be concerned only with the soliton part since it has the dominant amplitude.

The numerical results
show that some basic features of the KdV-equation are supported, but
illustrates also its shortcomings. As a test we first considered
a limit where effects of gravity were ignored and found propagation
of a moderate amplitude soliton shaped structure with a small 
damping. We then increased the gravitational acceleration term and found the damping to be 
counterbalanced at $G = 0.25$ resulting in a slow growth, and then for $G = 0.5$ we find an initially exponential growth that saturates for large times in qualitative agreement with the analytical predictions.

It is an essential element in the analysis that the linear energy propagation
speed (here the ion sound speed) is constant for all vertical positions, independent of density.
For a number of other wavetypes, also this speed is varying and the
energy density flux then becomes a competition between several parameters.
Phenomena and results similar to those studied here  can be found
for other inhomogeneous plasma conditions realizable in laboratory plasmas
\cite{doucet_jones_alexeff_1974,dangelo_michelsen_pecseli_1975,dangelo_michelsen_pecseli_1976,garcia_pecseli_2013,pecseli_2016}. We note though that plasma sheaths near solid
surface require models without assumptions of quasi neutrality. Such problems require a separate analysis. Conditions where a vertical flow is forced from $z=0$ in the direction opposed to gravity
is singular \cite{garcia_et_al_2015}, and requires a separate analysis.

\appendix*

\section{Boussinesq equations}

The KdV equation is explicitly derived for waves or pulses propagating in
one direction, as evidenced by the operator $\partial /\partial t
- C_s\partial /\partial z$ in the lowest order approximation. It is possible
to obtain an equation which can account for bi-directional propagation,
here given in dimensionless form \cite{drazin_johnson_1989}
\begin{equation}
\frac{\partial^2}{\partial t^2} u-\frac{\partial^2}{\partial z^2} u
-\frac{\partial^4}{\partial z^4}u +\frac{\partial^2}{\partial z^2} u^2
=0 \, .
\label{bussi}
\end{equation}
The two first terms correspond to the classical sound equation as might be
expected. The third term represents a dispersion, where we note that a term
like ${\partial^4}u/{\partial t^2 \partial z^2}$ might as well have been argued.
The last term represents the nonlinearity. The equation does not have any
significant advantage over the KdV equation, however, at least not as long
soliton dynamics is an issue. The point is that two counter-propagating
pulse overlap for only a small time, and do not manage to interact 
significantly. In case of {\em overtaking} interactions, the interaction
time is much longer, and the interaction becomes significant. This limit
is, however, well described by the KdV equation.

We can formulate a nonlinear equation that includes the Boussinesq equation
for homogeneous conditions and at the same time accounts for the
linear dispersion relation (\ref{simp_eq}) obtained for the gravitational 
inhomogeneous system. This modified equation has the form
\begin{equation}
\frac{\partial^2}{\partial t^2} u-\frac{\partial^2}{\partial z^2} u
-\frac{\partial^4}{\partial z^4}u +\frac{\partial^2}{\partial z^2} u^2
= g\frac{\partial}{\partial z} u, 
\label{gravitybussi}
\end{equation}
where $g$ is here a dimensionless measure of 
the gravitational acceleration. Equation (\ref{gravitybussi}) can 
be reduced to our modified KdV equation.

\bibliography{noter,drift,picsim,references}

\end{document}